\documentclass[12pt]{iopart}
\usepackage{array}
\expandafter\let\csname equation*\endcsname\relax
\expandafter\let\csname endequation*\endcsname\relax
\usepackage{graphicx}
\usepackage{mathtools}
\usepackage{amsmath}
\usepackage{amssymb}
\usepackage{iopams} 
\usepackage{bm}
\usepackage{graphicx}
\usepackage{xcolor}
\usepackage{epstopdf}
\usepackage{dcolumn}
\usepackage{bm}
\usepackage{amsfonts,amssymb,amsmath,array}
\usepackage{amsbsy}
\usepackage{color}
\usepackage{hyperref} 
\usepackage{enumerate}
\usepackage{mathtools}
\usepackage{lipsum} 

\usepackage{bigints}
\usepackage{comment}
\usepackage{array}

\usepackage{soul}

  \newcommand{\av}[1]{\left\langle#1\right\rangle}

  \newcommand{\new}[1]{#1}
  
  \newcommand{\out}[1]{}

\DeclareRobustCommand{\vect}[1]{
  \ifcat#1\relax
    \boldsymbol{#1}
  \else
    \mathbf{#1}
  \fi}

\newcommand{\lx} {\left}
\newcommand{\rx} {\right}
\newcommand{\eps} {\zeta}
\newcommand{\ener} {\varepsilon}
\newcommand{\Ener} {\mathcal{E}}

\begin{document}

\title{$H$-theorem at negative temperature: the random exchange model with bounds}

\author{Dario Lucente$^1,^2$, Marco Baldovin$^1$, Andrea Puglisi$^1$, Angelo Vulpiani$^3$}




\address{$^1$ Institute for Complex Systems, CNR, 00185, Rome, Italy}
\address{$^2$  Department of Mathematics \& Physics, University of Campania “Luigi Vanvitelli”, 81100, Caserta, Italy}
\address{$^3$ Dipartimento di Fisica, Universit\`a di Roma Sapienza, 
P.le Aldo Moro 5, 00185, Rome, Italy}
\ead{marco.baldovin@cnr.it}


\begin{abstract}
Random exchange kinetic models are widely employed to describe the conservative dynamics of large interacting systems. Due to their simplicity and generality, they are quite popular in several fields, from statistical mechanics to biophysics and economics. Here we study a version where bounds on the individual shares of the globally conserved quantity are introduced. We analytically show that this dynamics allows stationary states with population inversion, described by Boltzmann statistics at negative absolute temperature if the conserved quantity has the physical meaning of an energy. The proposed model provides therefore a privileged system for the study of thermalization toward a negative temperature state. First, the genuine equilibrium nature of the stationary state is verified by checking the detailed balance condition. Then, an $H$-theorem is proven, ensuring that such equilibrium condition is reached by a monotonic increase of  the Boltzmann entropy. We also provide analytical and numerical evidence that a large intruder in contact with the system thermalizes, suggesting a practical way to design a thermal bath at negative temperature. 
\end{abstract}

\section{Introduction} 

\new{In statistical physics it is not unusual to deal with systems whose elements, or particles, can only store a bounded amount of energy: nuclear spins~\cite{abragam1958spin} and vortexes in 2-dimensional hydrodynamics~\cite{Onsager1949} are important examples. When such systems are isolated from the environment, and their internal energy is large enough, inverted-population states are known to occur, characterized by an ordered phase which is qualitatively different from that of the ground state. For instance, at very large energy a spin lattice with ferromagnetic interactions shows an antiferromagnetic phase~\cite{hakonen92}, while point vortexes in a confined domain experience clusterization~\cite{Gauthier2019,Johnstone2019}. Indeed, after a certain threshold, the volume of available phase space (and hence entropy) decreases when the internal energy increases: the larger the energy, the more ordered the phase. Consistently, these states are well described by a Boltzmann statistics with negative inverse temperature $\beta$.} Such \textit{negative absolute temperature} states were first studied by Onsager for 2-dimensional hydrodynamics~\cite{Onsager1949} and then experimentally observed in nuclear spin systems~\cite{Purcell1951,Ramsey1956, hakonen92}, cold atoms~\cite{Braun2013} and superfluids~\cite{Gauthier2019,Johnstone2019}. They also naturally emerge in the study of the Nonlinear Discrete Shr\"{o}dinger Equation~\cite{Rasmussen2000,Iubini2012,Iubini2017,Gradenigo2021}. \new{These states are instead forbidden whenever the single-particle Hamiltonian includes a kinetic contribution in the usual form, quadratic in the momentum: if this is the case, single-particle energy is no longer bounded, and there is no way to get a decreasing dependence of the entropy on energy~\cite{cerino2015consistent,baldovin2017thermometers,  Baldovin2021}. Nonetheless, negative temperature can be observed for kinetic degrees of freedom if the Hamiltonian has a different dependence on the momentum~\cite{Braun2013, cerino2015consistent, Baldovin2018, Baldovin2019}. Alternatively, it would be possible to have transient negative temperature states if the kinetic part of the Hamiltonian relaxed on a time scale much longer than that of the potential.}

Whether states at negative absolute temperatures are, or are not, genuine equilibrium states has been the topic of a long debate in the statistical mechanics community~\cite{Dunkel2013, Buonsante2016, Struchtrup2018, cerino2015consistent, Swendsen2018,  Baldovin2021}. \new{Part of the confusion comes from the fact that, since one deals with isolated systems, ``equilibrium'' refers here to the internal condition of the system, rather than its relative state with the environment. This point is made clear, for instance, in the seminal paper~\cite{Purcell1951}, where the system is observed to equilibrate to a negative-temperature state, but only on time scales much shorter than the typical relaxation with the surrounding environment. However, it is argued in~\cite{Struchtrup2018} that  negative-temperature stationary distributions should be \textit{always} regarded as peculiar out-of-equilibrium states, even when the system is isolated from the environment.} The claim is motivated by the Kelvin-Planck version of the Second Law of Thermodynamics, which in its original formulation (conceived, of course, for positive-temperature systems) would lead to inconsistencies in the presence of equilibrium at negative temperature. The issue was already known to Ramsey, who proposed a generalization of the Kelvin-Planck formulation to systems that can allow negative temperature~\cite{Ramsey1956}; \new{in~\cite{Struchtrup2018} it is proposed instead to refuse this fixing, by completely renouncing the concept of negative-temperature equilibrium (and therefore all the results that can be obtained for bounded-energy systems in the framework of equilibrium statistical mechanics~\cite{Baldovin2021}). In this work, we discuss an example that shows, beyond doubt, that stationary states at negative temperature should be regarded as equilibrium states, supporting Ramsey's point of view.}


\new{The system we are going to consider belongs to the class of random exchange models.}
Most  processes in statistical physics are characterized by several units experiencing pairwise interactions that conserve some fundamental quantities, typically energy and momentum. The dynamical evolution of these systems can be described by kinetic models, where collisions are treated statistically. Of course the most famous one is the Boltzmann equation for rarefied gases, other noticeable examples being represented by the Landau equation for collisional plasma and the Fokker-Planck equation for the dynamics of colloidal particles in contact with a thermal bath~\cite{villani2002review}.
In some cases it is useful to reduce the process to its essential ingredients, that is random collisions with energy conservation, without taking into account the details of the interaction. This is the idea behind the Boltzmann-like model proposed by Ulam and Everett in~\cite{everett1969entropy} and later developed in~\cite{ulam1980operations}, where the sum of two colliding particles is randomly distributed between them according to a given distribution $p(\alpha)$. Motivated \textcolor{black}{both by the theoretical findings of ~\cite{everett1969entropy}, where a class of nonlinear dynamics having the exponential distribution as steady state and the Boltzmann entropy as Lyapunov function is introduced, and} by some numerical results \textcolor{black}{on the general case, Ulam}  
conjectured that this process admits a stationary state (the proof was provided a few years later~\cite{blackwell1985ulam}). 
\textcolor{black}{His} interest in these models arose from their potential usefulness in biological processes, as extensively documented in~\cite{ulam1980operations}. His idea is strictly connected with a class of  kinetic models for  molecules that interact through a particular power-law decaying, soft repulsion potential, the so-called Maxwell model~\cite{ernst1981nonlinear}.
A variant of this dynamics with inelastic collisions  has also been extensively studied in the literature of granular kinetic models~\cite{baldassarri2002kinetics}. 
Over the last half century, simplified versions have been used in several contexts, from biology (where they are used to model cells mutation processes~\cite{pareschi2023kinetic} or the velocity exchange dynamics in flocks~\cite{carrillo2010asymptotic,cao2020asymptotic}) to financial applications (see~\cite{dragulescu2000statistical,chakraborti2000statistical,chatterjee2003money,greenberg2023twenty} and references therein), \textcolor{black}{as well as for studying transport phenomena, condensation and jamming~\cite{zia2004construction,majumdar2005nature,bertin2005subdiffusion,bertin2006definition,guioth2017mass}}. 
In the seminal paper~\cite{dragulescu2000statistical}, the authors replaced the energy distribution with distributions of the wealth in closed economic systems. 
Since then, as well reported in the recent review~\cite{greenberg2023twenty}, various authors studied slightly modified exchange models borrowing concept from statistical mechanics. For instances, in~\cite{boghosian2017oligarchy} it is shown that a class of random exchange models introduced in~\cite{boghosian2014kinetics} displays a second order phase transition. Although the interest in these systems arises mainly from the properties of the steady state, some authors studied how this steady state is approached. Notably, in a series of papers~\cite{apenko2013monotonic,apenko2014clausius}, Apenko, exploiting the formulation of the problem introduced in~\cite{lopez2012exponential}, was able to prove that the Boltzmann entropy is a Lyapunov functional for the original Everett-Ulam model. A similar result was obtained in~\cite{boghosian2015h}, where the authors showed that the Gini coefficient (a measure which quantifies the degree of inequalities in a given wealth distribution) acts as a Lyapunov functional for their dynamics.
\\
\indent In this paper we study a modified version of the random exchange model, where the internal energy of each element has a given upper bound: after each collision, the energy of the pair is randomly redistributed between the two particles in such a way that none of the two exceeds that threshold. 
 The considered random exchange model with bounds admits stationary states at negative temperature. It represents an ideal case study for the understanding of the statistical properties of these states, and in particular for showing their genuine equilibrium nature. To support this point, \new{not only we prove} that detailed balance holds for a particle's dynamics in a system at stationarity, \new{but we also} exhibit a rigorous proof of an $H$-theorem for this dynamics, inspired by the works of Apenko~\cite{apenko2013monotonic,apenko2014clausius}, showing that the stationary distribution is monotonically reached.
Numerical simulations of systems with a large-energy intruder further show that the random exchange model can act as a thermal bath at negative absolute temperature.  

The paper is organized as follows. In Section~\ref{sec:model} we introduce the nonlinear Boltzmann-like equation that governs the temporal evolution of the one-particle probability density function $\rho_t$, providing also the expression of the stationary solution of the model. Section~\ref{sec:H-theorem} is devoted to the analytical derivation of an $H$-theorem for this process. Its equilibrium nature is proved in Section~\ref{sec:probe_particle}, where it is shown that the detailed balance condition holds for the dynamics of a probe particle. \new{We also discuss} the use of the model as a thermal bath: the thermalization of a massive particle towards negative temperature states is explicitly analyzed.

 \section{Model}\label{sec:model} 
Let us consider a system composed of $N$ particles, characterised by non-negative energy values $\Ener_n(t)<\Ener_M$, with $1\le n \le N$, where $\Ener_M$ is a given upper bound. The evolution follows a discrete-time dynamics. At each time step, two particles $i$ and $j$ are randomly chosen, and they exchange a certain amount of energy in such a way that $\Ener_i(t+1)$ and $\Ener_j(t+1)$ are still lower than $\Ener_M$. In formulae:
\begin{equation}
\begin{aligned}
&\Ener_i(t+1) = \alpha \lx(\Ener_i(t)+\Ener_j(t)\rx)+\Delta_{ij}(t),\\
&\Ener_j(t+1) = \lx(1- \alpha \rx) \lx(\Ener_i(t)+\Ener_j(t)\rx)-\Delta_{ij}(t)\,,
\label{eq:evolution_two_particle_neg}
\end{aligned}
\end{equation}
with
$$
\Delta_{ij}(t)=\Ener_M\lx(1-2\alpha\rx)\theta\lx(\Ener_i(t)+\Ener_j(t)-\Ener_{M}\rx)\,.
$$
Here $\theta(x)$ is the Heavyside step-function and $\alpha \in[0,1]$ is extracted at each time step according to some probability density function. In the following we will limit the discussion to the uniform distribution
$$
p( \alpha )=\theta(\alpha)\theta(1-\alpha)\,.
$$
Note that $\Delta_{ij}(t)$ is different from $0$ only if $\Ener_i+\Ener_j>\Ener_M$. The evolution introduced above consists in redistributing the energies $\{\Ener_n\}$ or the ``vacancies'' $\{\Ener_M-\Ener_n\}$ depending on the sign of $\Ener_M-\Ener_i-\Ener_j$, in such a way that all outcomes compatible with the energy bounds are equally probable.  \textcolor{black}{At variance with the kinetically constrained model proposed in~\cite{bertin2005subdiffusion}, where energies (interpreted as masses) are redistributed only if their sum does not exceed a given threshold, the dynamical rule introduced here prevents the appearance of jammed states and it is therefore more suitable for studying population inversion phenomena.}

If $\Ener_M$ is larger than the total energy of the system, $\Ener_M > \sum_n \Ener_n$,  the term $\Delta_{ij}$ identically vanishes and one recovers the original model by Ulam~\cite{ulam1980operations}. In this case, when the thermodynamic limit ($N\to\infty$) is considered, the single particle distribution $\rho_t(\Ener)$ follows a nonlinear evolution, which is a discrete time version of the Boltzmann equation~\cite{ulam1980operations,mauldin1987mathematical,cao2023entropy}. Furthermore, it has been proved that it reaches a stationary state described by the equilibrium distribution~\cite{blackwell1985ulam} 
\begin{equation}
\rho_{\infty}(\Ener)=\beta\exp(-\beta \Ener)\,.    
\end{equation}
By introducing an auxiliary linear process in two dimensions (the pair space), it is possible to show that the Boltzmann entropy
$$
S_B[\rho_t]=-\int {\rm d}\Ener\,\rho_t(\Ener)\log \rho_t(\Ener)
$$
is a Lyapunov function, i.e., it grows monotonically during the evolution~\cite{apenko2013monotonic,apenko2014clausius}. 

 If $\Ener_M<\sum_n \Ener_n$, the $\Delta_{ij}$ contributions on the rhs of Eq.~\eqref{eq:evolution_two_particle_neg} are different from zero. An evolution of this sort is expected, for instance, when considering a system of isolated nuclear spins $\{\sigma_n\}$ immersed in a strong magnetic field $B$: such dynamics is due to the sudden exchanges of energy within spin pairs, with the constraint that each of the $\{\Ener_n\}$ is bounded by $2 B \sigma_M $ (here $\sigma_M$ is the maximum value achievable by the spin)~\cite{Ramsey1956,abragam1958spin, oja97}. Equation~\eqref{eq:evolution_two_particle_neg} can be seen as the limit of this dynamics when the spins have continuous values (as in the $XY$ model). 

The notation can be simplified by performing the change of variables
\begin{subequations}
\begin{align}
    \ener_i&=\frac{2\Ener_i-\Ener_M}{2\Ener_M}\,,\\
    u&=\lx(2 \alpha -1\rx)\,,\\
     \eps_{ij}&=\lx(\ener_i+\ener_j\rx)/2\,,\\
    f(\eps)&=\eps-\frac{1}{2}\text{sign}(\eps)\,.
\end{align}    
\end{subequations}
With the above definitions, $\ener_i$, $\eps_{ij}$ and $f(\eps_{ij})$ assume values in the interval $[-1/2,1/2]$, while $u \in [-1,1]$.
We rewrite Eq.~\eqref{eq:evolution_two_particle_neg} as
\begin{equation}
\begin{aligned}
   \ener_i(t+1) = \eps_{ij} + u f(\eps_{ij})\,,\\
   \ener_j(t+1) = \eps_{ij} - u f(\eps_{ij})\,.
\end{aligned}    
\end{equation}
The evolution law of the single-particle energy distribution $\rho_t(\ener)$ can be written as
\begin{align}
    \rho_{t+1}(\ener)&=\int { {\rm d} u\, {\rm d}\ener_{1} \,{\rm d}\ener_{2} \,p\lx(u\rx)\rho_t(\ener_{1})\rho_t(\ener_{2})\delta[\ener-\eps_{12}-uf(\eps_{12})]}\nonumber\\
    &=\int{\rm d}\ener_{1}\,{\rm d}\ener_{2} \,p\lx(\frac{\ener-\eps_{12}}{f\lx(\eps_{12}\rx)}\rx)\frac{\rho_t(\ener_{1})\rho_t(\ener_{2})}{|f\lx(\eps_{12}\rx)|}
\label{eq:Ulam_evolution_pdf_negative}
\end{align} 
where 
\begin{equation}
   p(u)=\frac{1}{2}\theta(u+1)\theta(1-u)\,.
   \label{eq:distributionlaw}
\end{equation}
 An explicit check shows that 
\begin{equation}
\rho_{\infty}\lx( \ener \rx)=\frac{\beta}{2\sinh\lx(\frac{\beta}{2}\rx)} \exp\lx(-\beta \ener\rx)
\label{eq:asymptotic_distribution}
\end{equation}
is a fixed point of 
\eqref{eq:Ulam_evolution_pdf_negative}.  
The main panel of Fig.~\ref{fig:EntropyVsTime_Distribution}, showing the evolution of $\rho_t$ according to a numerical integration of Eq.~\eqref{eq:Ulam_evolution_pdf_negative}, confirms that $\rho_{\infty}\lx( \ener \rx)$ is asymptotically reached starting from atypical conditions. The parameter $\beta$ - which fixes the mean energy per particle $\av{\ener}$ - can take both positive and negative values, depending on the sign of $\av{\ener}$. To determine its value it is sufficient to solve (numerically) the transcendental equation $$
\beta\av{\ener}=1-\frac{\beta}{2}\coth\lx(\frac{\beta}{2}\rx)\,.
$$

\section{$H$-theorem}\label{sec:H-theorem}

Despite the differences with the case originally studied by Ulam, also for the bounded-energy version of the model considered here it is possible to prove an $H$-theorem, following a strategy inspired by~\cite{apenko2013monotonic,apenko2014clausius}. \textcolor{black}{
Before going into the details of the demonstration, we believe it is useful to highlight the difficulties that prevent a simple derivation directly from the $N-$particles dynamics~\eqref{eq:evolution_two_particle_neg}. 
Since this dynamics is Markovian and ergodic, the process~\eqref{eq:evolution_two_particle_neg} admits a unique stationary solution $\mathcal{P}_\infty^{(N)}\left(\Ener_1,\cdots,\Ener_N\right)$ while the Kullback-Leibler divergence $K\left(\mathcal{P}_t^{(N)}||\mathcal{P}_\infty^{(N)}\right)$ decreases monotonically during the evolution~\cite{van1992stochastic}.
Moreover, when the uniform distribution for $p(\alpha)$ is considered, the $N-$body distribution $\mathcal{P}_\infty^{(N)}\left(\Ener_1,\cdots,\Ener_N\right)$ coincides with the microcanonical distribution\footnote{More generally, it is possible to prove that if $p(\alpha)$ is a Beta distribution then  $\mathcal{P}_\infty^{(N)}\left(\Ener_1,\cdots,\Ener_N\right)=\frac{1}{Z_N}\prod_{i=1}^N f(\Ener_i)\delta\left(\Ener-\sum_{n}\Ener_n\right)$ where $Z_N$ is a normalization constant/ the partition function~\cite{zia2004construction,bertin2005subdiffusion,bertin2006definition}}, i.e.
\begin{equation}
	\mathcal{P}_\infty^{(N)}\left(\Ener_1,\cdots,\Ener_N\right)=\frac{1}{\mathcal{N}}\,,
\end{equation}
where $\mathcal{N}$ is the number of microstates with total energy $\mathcal{E}=\sum_n\Ener_n$. Thus, the relative entropy $K$ boils down to the Shannon entropy $S_B[\mathcal{P}_t^{(N)}]$. However, for any fixed $N$, both the global energy constraint, $\mathcal{E}=\sum_n\Ener_n$, as well as the local energy conservation involved in pairwise collision, $\Ener_i(t+1)+\Ener_j(t+1)=\Ener_i(t)+\Ener_j(t)$, induce correlation on the system: in general $\mathcal{P}_t^{(N)}$ is not factorised and $S_B[\mathcal{P}_t^{(N)}]\neq NS_B[\mathcal{P}_t^{(1)}]$ (where $\mathcal{P}_t^{(1)}\equiv\rho_t$). Hence, the monotonic increase of $S_B[\rho_t]$ along the evolution has to be proven for the nonlinear dynamics~\eqref{eq:Ulam_evolution_pdf_negative} which represents the mean-field equation governing the evolution of the system in the thermodynamic limit. } 
 To this end, let us consider the linear transformation 
\begin{equation}
    \eta_{t+1}(\ener_1,\ener_2)=\int_{-1}^1{\rm d}u\,\, \frac{1}{2} \eta_t\lx(\eps_{12} + u f(\eps_{12}),\eps_{12} - u f(\eps_{12})\rx)\,,
\label{eq:linear_dynamics}
\end{equation} 
which describes the evolution of the probability density function (pdf) of a pair of particles. This equation admits infinitely many fixed points~\cite{apenko2013monotonic,apenko2014clausius}, including the exponential distribution 
$$
\eta_{\infty}(\ener_1,\ener_2)=\frac{\beta^2  \exp\lx(-\beta\lx(\ener_1+\ener_2\rx)\rx)}{2\lx(\cosh\beta-1\rx)}\,.
$$
It is useful to introduce
$$
\mu_t(u,\ener_1,\ener_2)=\frac{1}{2}\eta_t\lx(\eps + u f(\eps),\eps - u f(\eps)\rx)\,,\\ 
$$
and to denote by $\tilde{\mu}_{t}$ 
its marginal
$$
\tilde{\mu}_{t}(\ener_1,\ener_2)=\int_{-1}^1 du \, \mu_{t}(u,\ener_1,\ener_2)\,.
$$
Note that with this definition
$$
\tilde{\mu}_{t}=\eta_{t+1}\,.
$$

A general result in information theory~\cite{Cover2006} guarantees that a coarse-grained procedure necessarily reduces the relative entropy (Kullback-Leibler divergence)~\cite{esposito2012stochastic}, i.e.  
\begin{equation}
    K(\mu_t||\mu_{\infty})\ge K(\tilde{\mu}_{t}||\tilde{\mu}_{\infty})\,,
    \label{eq:bound_relative_entropy}
\end{equation}
where
\begin{align}
    K(\mu_t||\mu_{\infty})&=\int {\rm d}u\,{\rm d}\ener_1\,{\rm d}\ener_2\,\mu_t(u,\ener_1,\ener_2)\log\lx(\frac{\mu_t(u,\ener_1,\ener_2)}{\mu_{\infty}(u,\ener_1,\ener_2)}\rx)\,,\label{eq:kullback}\\
    K(\tilde{\mu}_{t}||\tilde{\mu}_{\infty})&=\int {\rm d}\ener_1\,{\rm d}\ener_2\,\tilde{\mu}_{t}(\ener_1,\ener_2)\log\lx(\frac{\tilde{\mu}_{t}(\ener_1,\ener_2)}{\tilde{\mu}_{\infty}(\ener_1,\ener_2)}\rx)\nonumber\\
    &=\int {\rm d}\ener_1\,{\rm d}\ener_2\,\eta_{t+1}(\ener_1,\ener_2)\log\lx(\frac{\eta_{t+1}(\ener_1,\ener_2)}{\eta_{\infty}(\ener_1,\ener_2)}\rx)\,.\label{eq:kullback_cg}
\end{align}
Performing the change of variables
$$
\begin{aligned}
x&=\eps_{12}+uf(\eps_{12})\\
y&=\eps_{12}-uf(\eps_{12})\\  z&=\ener_1-\ener_2    
\end{aligned}
$$
on Eq.~\eqref{eq:kullback}, one obtains \begin{equation}
K(\mu_t||\mu_{\infty})= \int_{-\frac{1}{2}}^{\frac{1}{2}}{\rm d}x\,\int_{-\frac{1}{2}}^{\frac{1}{2}}{\rm d} y \,\,\eta_t(x,y) \log\lx(\frac{\eta_t(x,y)}{\eta_\infty(x,y)}\rx)\,.
\label{eq:kullback_change_variable}
\end{equation}
Thus, combining Eq.~\eqref{eq:bound_relative_entropy} and Eq.~\eqref{eq:kullback_change_variable}, we prove that $K(\eta_{t}||\eta_{\infty})\ge K(\eta_{t+1}||\eta_{\infty})$, \textcolor{black}{that is 
\begin{equation}
	\langle \log(\eta_{t})\rangle_{\eta_t}-\langle\log(\eta_{\infty})\rangle_{\eta_t} \ge \langle\log(\eta_{t+1})\rangle_{\eta_{t+1}}-\langle\log(\eta_{\infty})\rangle_{\eta_{t+1}}\,.
\end{equation}
Since $\eta_{\infty}$ is a function of $x+y$ and the linear transformation~\eqref{eq:linear_dynamics} preserves the sum, one has \begin{equation}
\langle\log(\eta_{\infty})\rangle_{\eta_t}=-\beta\langle(\ener_1+\ener_2)\rangle+\log\lx(\frac{\beta^2}{2\lx(\cosh\beta-1\rx)}\rx)= \langle\log(\eta_{\infty})\rangle_{\eta_{t+1}}
\end{equation}
and the above equation implies}
\begin{equation}
S_B(\eta_{t+1})\ge S_B(\eta_{t})\,.
\label{eq:$H$-theorem_2d_linear}
\end{equation}
The monotonic behaviour of the relative entropy $K(\eta_{t}||\eta_{\infty})$ could have been deduced as well on the basis of general properties of Markov processes~\cite{van1992stochastic}; the same is not true for the Boltzmann entropy $S_B(\eta_t)$.

Equation~\eqref{eq:$H$-theorem_2d_linear} can be used to prove an $H$-theorem for the one-particle pdf $\rho_t(\ener)$.
First, let us notice that by defining 
$$
\eta_t(x,y)\equiv \rho_t(x)\rho_t(y)\,,
$$
the quantity $\eta_{t+1}(x,y)$ given by Eq.~\eqref{eq:linear_dynamics} fulfills
\begin{equation}
    \rho_{t+1}(x)=\int {\rm d}y\, \eta_{t+1}(x,y)\,.
    \label{eq:marginalization_two_particle}
\end{equation}
In other words, Eqs.~\eqref{eq:marginalization_two_particle} and \eqref{eq:linear_dynamics} are equivalent to the evolution \eqref{eq:Ulam_evolution_pdf_negative}. 
Since $\rho_{t+1}$ is the marginal of $\eta_{t+1}$, one can use a known result from information theory to prove the $H-$theorem. Indeed the mutual information between $\eta_{t+1}(x,y)$  and $\rho_{t+1}(x)\rho_{t+1}(y)$ (which is non-negative by definition) verifies~\cite{Cover2006}
\begin{equation}
\begin{split}
    I&=\int \int {\rm d}y\, {\rm d}x\,\eta_{t+1}(x,y)\log\lx(\frac{\eta_{t+1}(x,y)}{\rho_{t+1}(x)\rho_{t+1}(y)}\rx)=\\
    &=2S_B(\rho_{t+1})-S_B(\eta_{t+1})\ge 0\,,
\end{split}
\end{equation}
which combined with Eq.~\eqref{eq:$H$-theorem_2d_linear} yields
\begin{equation}
S_B(\rho_{t+1})\ge\frac{1}{2}S_B(\eta_{t+1})\ge \frac{1}{2}S_B(\eta_{t})=S_B(\rho_t)\,.
\end{equation}

\begin{figure}
\centering    \includegraphics[width=.8\linewidth]{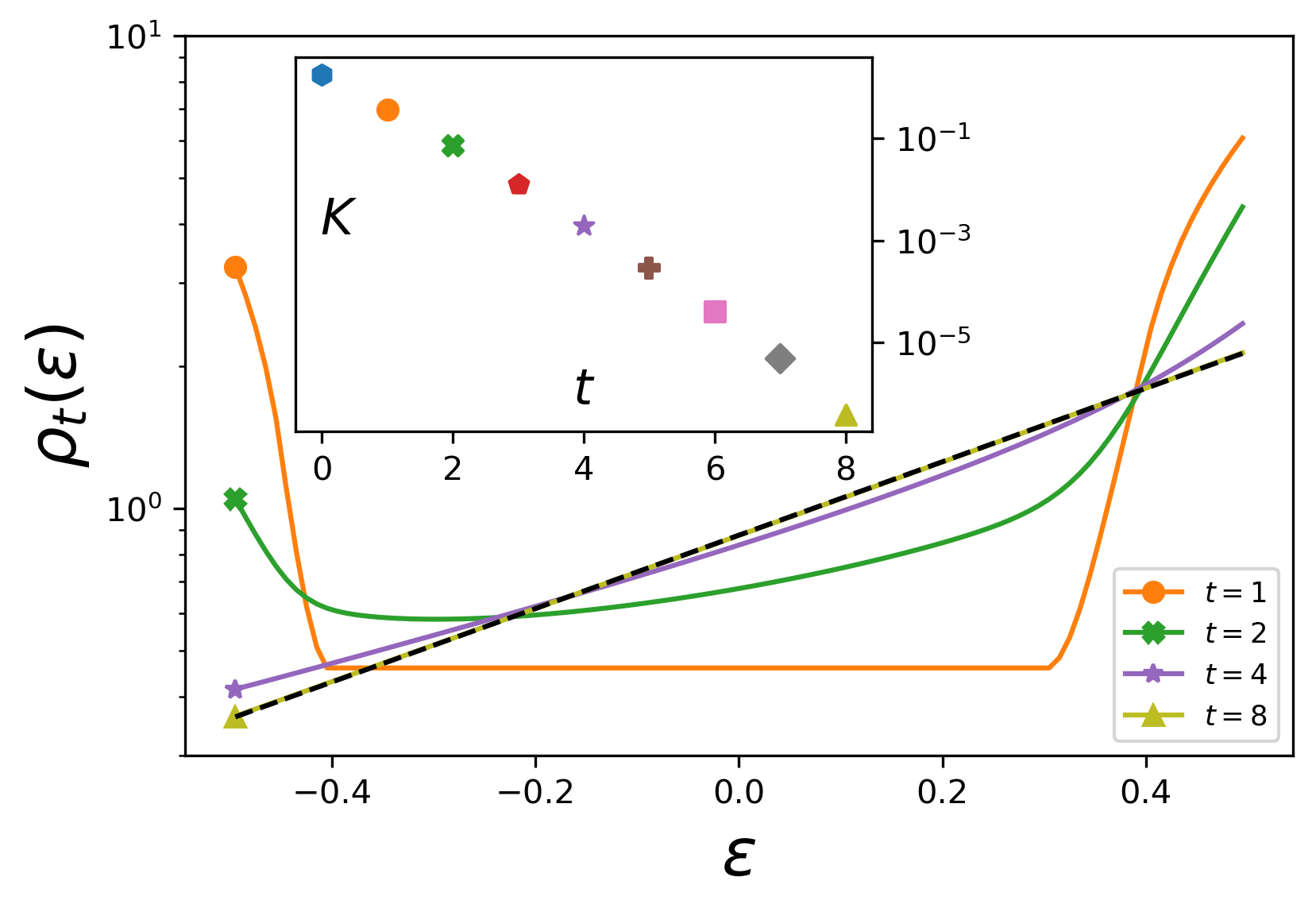}
 \caption{Convergence of the single-particle pdf $\rho_t(\ener)$ towards the asymptotic exponential distribution (dashed black line) starting from a uniform distribution over the disconnected domain $[-0.5,-0.45]\cup[0.4,0.5]$.
 The symbols on the curves (as well as their colors) are in one-to-one correspondence with those in the inset. Inset: time evolution of the relative entropy $K$. 
The parameters are $\av{\ener}\approx 0.1416$ and $\beta\approx -1.789$.
 }
  \label{fig:EntropyVsTime_Distribution}
\end{figure}

The monotonic behavior of the relative entropy $K$ is shown on the inset of Fig.~\ref{fig:EntropyVsTime_Distribution}. It should be noted that the convergence towards the asymptotic distribution is exponentially fast.

\section{Dynamics of a probe particle in the stationary state}\label{sec:probe_particle}
 
We now focus our attention on the evolution of the pdf $\pi_t(\ener)$ of a probe particle interacting with $N\gg 1$ particles whose energies are distributed according to $\rho_\infty(\ener)$.
\new{Our aim is to provide clear evidence of the genuine equilibrium nature of stationary states with negative temperature. As we will explain better shortly, in the steady state the dynamics is equivalent to a Markovian one. Then, proving equilibrium consists in showing that detailed balance holds. Moreover, once equilibrium is established, it becomes natural to investigate whether the system can be used as a thermal bath. To this aim, considering a "massive" probe particle, we show that the system actually acts as a thermostat at both positive and negative temperatures.}

\subsection{\new{Equilibrium nature of the stationary state}}
The evolution of $\pi_t(\ener)$ can be conveniently written
as
\begin{equation}
    \pi_{t+1}(\ener)=\int{ {\rm d}\ener_{1} } \pi_t(\ener_{1}) W(\ener_1\to \ener)
\end{equation}
where the transition rate $W$ is given by
\begin{equation}
    W(\ener_1\to \ener)=\int{{\rm d}\eps \,p\lx(\frac{\ener-\eps}{f\lx(\eps\rx)}\rx)\frac{2b(\eps,\ener_1)\rho_\infty(2\eps-\ener_1)}{|f\lx(\eps\rx)|}}\,,
\end{equation}
and the function $b(\eps,\ener_1)$ enforces the integration domain:
\begin{equation}
b(\eps,\ener_1)=\theta\lx(2\eps-\ener_1+1/2\rx)\theta\lx(\ener_1+1/2-2\eps\rx)\theta\lx(\ener_1+1/2\rx)\theta\lx(1/2-\ener_1\rx)\,.    
\end{equation}
Thus, the evolution of $\pi_t(\ener)$ is Markovian and $\lim_{t\to\infty}\pi_t(\ener)=\rho_{\infty}(\ener)$.
Recalling~\eqref{eq:distributionlaw}, we get
\begin{equation}
p\lx(\frac{\ener-\eps}{f\lx(\eps\rx)}\rx)=\frac{1}{2}b\lx( \eps, \ener\rx)\,.
\label{eq:pbdetailed}
\end{equation}
Besides, since $\rho_{\infty}(\ener)$ is exponential, one has $\rho_\infty(\ener_1)\rho_\infty(2\eps-\ener_1)\propto\rho_\infty(2\eps)$. As a consequence,
$$
\rho_\infty(\ener_1)W(\ener_1\to \ener)=\rho_\infty(\ener)W(\ener\to\ener_1)
$$
i.e. the detailed balance condition  is always verified, meaning that the dynamics is invariant under time-reversal and the system is at equilibrium~\cite{gardiner85}.


\subsection{A thermal bath at negative temperature}\label{sec:thermal_bath}
Previous studies of a thermal bath at negative temperature were performed in the context of two-dimensional hydrodynamics~\cite{Chavanis1998} and for Hamiltonian systems in contact  with deterministic~\cite{Baldovin2018} and stochastic reservoirs~\cite{Baldovin2019}. Monte Carlo numerical schemes have been devised in order to simulate the presence of a negative temperature bath~\cite{Iubini2012, Iubini2017}.
Here we show that the modified Ulam model is particularly suitable for being used as a thermostat both at positive and negative temperature. 

Consider a particle whose energy $E$ is between $\lx[0,E_M\rx]$ surrounded by a very large number $N$ (ideally infinite) of particles with energies $\Ener\in\lx[0,\Ener_M\rx]$ with $\Ener_M\ll E_M$. The particles belonging to the thermal bath can exchange energy with each other through the collisional rule~\eqref{eq:evolution_two_particle_neg}. Furthermore, these can also exchange energy with the intruder. Since the number of particles in the thermal bath is very large and the particle energies are small compared with $E_M$, the time required to reach equilibrium will be much less than that of the intruder. Therefore we can assume the energies of the bath particles to be distributed as $\rho_{\infty}(\Ener)\propto \exp\lx(-\beta \Ener\rx)$ and  focus on the evolution of the probability distribution $\mathcal{P}_t(E)$ of the intruder. Since  $\mathcal{P}_t(E)$ and $\rho_{\infty}(\Ener)$ have different supports, the collisional rule described in the previous section is not directly applicable. However, as already anticipated, that rule is equivalent to requiring that all possible outcomes have the same probability. Thus, in the same spirit, we can define the collision as follows
\begin{align}
E(t+1) = E(t)-\Delta(E,\Ener) \,,\nonumber\\
\Ener(t+1) = \Ener(t)+\Delta(E,\Ener)\,,
\label{eq:evolution_intruder_particle}
\end{align}
with $\Delta\in\{-\Delta_m,\Delta_M\}$, $\Delta_m=\text{min}\{\Ener,E_M-E\}$, $\Delta_M=\text{min}\{\Ener_M-\Ener,E\}$ and $$g(\Delta|\Ener,E)=\frac{\theta(\Delta+\Delta_m)\theta(\Delta_M-\Delta)}{\Delta_M+\Delta_m}\,.$$ 

\begin{figure}[ht]
\centering
    \includegraphics[width=.8\linewidth]{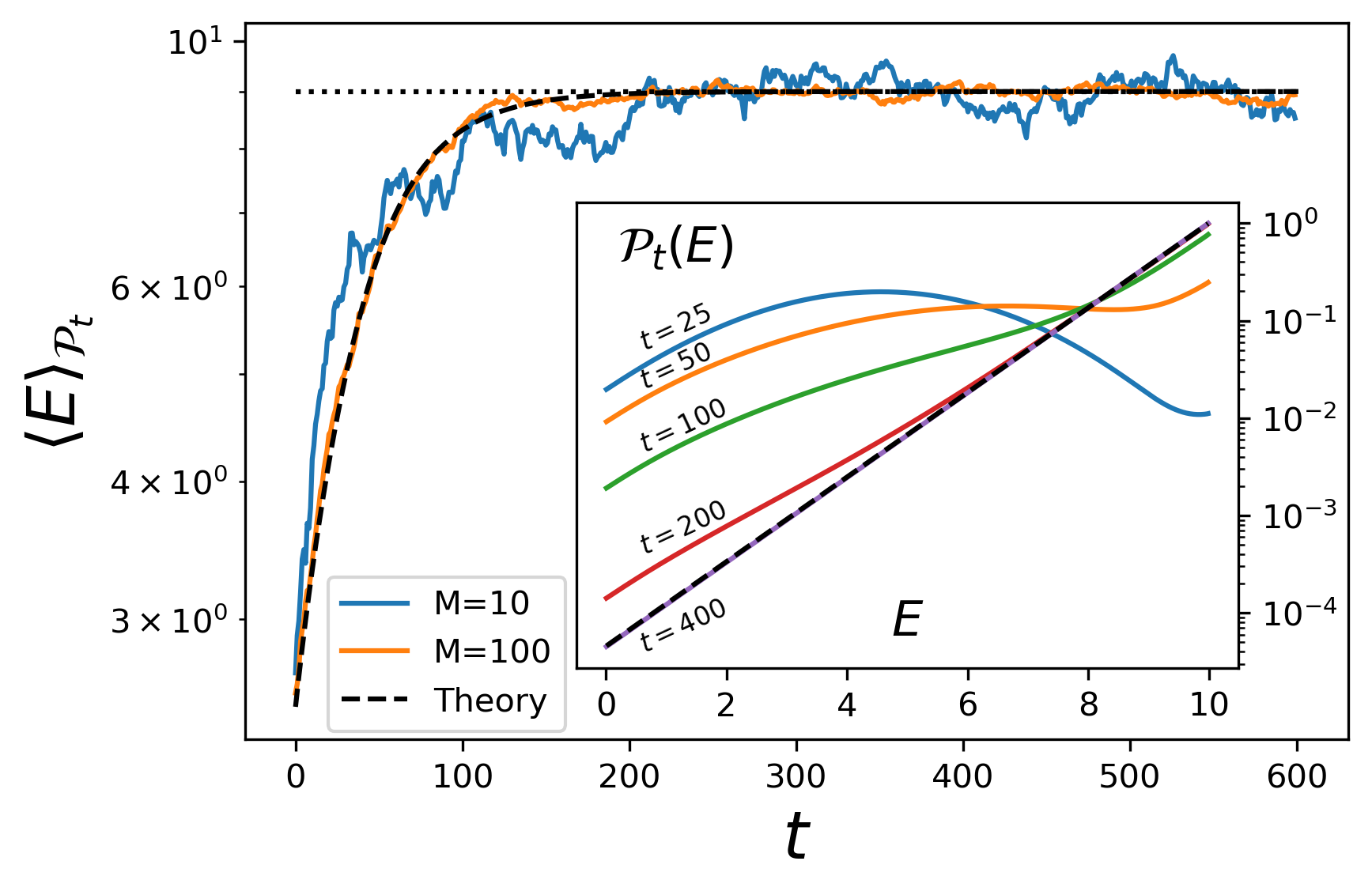}
 \caption{Mean energy of the massive intruder as a function of time obtained both through numerical integration of Eq.~\eqref{eq:Linear_evolution_intruder} (dashed black line) and as an average performed over $M$ realizations of numerical simulations of Eqs.~\eqref{eq:evolution_intruder_particle} 
 ($M=10$ is represented with a blue curve while the orange one represents $M=100$). The symbols correspond to the mean values of the pdfs shown in the inset. The parameters are $\Ener_M=1$, $E_M=10$, $\beta=1.$ and the number of light particles in the simulation is $N=10^4$. Inset: Convergence of the single-particle distribution $\mathcal{P}_t(E)$ towards the asymptotic exponential pdf (dashed black line)}
  \label{fig:MeanEnergyVsTime}
\end{figure}

Thus, the evolution of the pdf of the intruder is linear and Markovian, and in the limit of $N\gg 1$, $E_M\gg \Ener_M$ takes the form
\begin{align}
\mathcal{P}_{t+1}(E)&=\int {\rm d}E'\,{\rm d}\Ener\,\mathcal{P}_{t}(E')g(E'-E|E',\Ener)\rho_{\infty}(e)=\nonumber\\
&=\int {\rm d}E'\,\mathcal{P}_{t}(E')W(E'\to E)
\label{eq:Linear_evolution_intruder}
\end{align}
where $W(E'\to E)=\int {\rm d}\Ener\,g(E'-E|E',\Ener)\rho_{\infty}(\Ener)$.
The Markovian evolution guarantees that a limiting pdf $\mathcal{P}_{\infty}(E)$ exists and that the relative entropy \new{$K$} decreases monotonically. Furthermore, it can be verified that 
\begin{equation}
    \mathcal{P}_\infty(E)=\frac{\beta}{1-\exp\lx(-\beta E_M\rx)}\exp\lx(-\beta E\rx)
    \label{eq:asymptotic_distribution_intruder}
\end{equation}
is a fixed point of transformation~\eqref{eq:Linear_evolution_intruder} (see also the inset of Fig.~\ref{fig:MeanEnergyVsTime}). This means that the system of $N$ light particles can be regarded as a genuine thermal bath, given that at thermal equilibrium the massive intruder has the same temperature $1/\beta$. It should be noted, however, that since the energies are not quadratic functions of momenta, the equipartition theorem does not hold at thermal equilibrium. Notwithstanding, since the average energy $\av{E}$ is a function $\beta$ it is easily predictable knowing the bath temperature. Fig.~\ref{fig:MeanEnergyVsTime} shows the average energy of the massive intruder as a function of time obtained by numerical integration of Eq.~\eqref{eq:Linear_evolution_intruder} (black dashed line)  as well as the average performed over $M$ realizations (blue and orange curves) obtained by a direct simulation of Eqs.~\eqref{eq:evolution_intruder_particle} where a massive intruder collides with the $N$ light particles of the thermal baths.
At $t=0$, the intruder is in a state characterized by a positive temperature $\av{E}<\frac{E_M}{2}$ while the temperature of the thermal bath is negative ($\beta=-1$). As the intruder collides with the particles of the thermal bath, it acquires energy and its temperature decreases until it reaches the stationary state characterized by the exponential distribution (Eq.~\eqref{eq:asymptotic_distribution_intruder}). 
Again, it can be noted that the convergence is exponentially fast and its characteristic time scale $\tau$ is related to the ratio of the bounds, that is $\tau=O(\frac{E_M}{\Ener_M})$. Interestingly, these observations are valid also for small ensembles ($M=10$) up to fluctuations of order $1/\sqrt{M}$. Already with $M=100$ the differences between the numerical simulation and the numerical integration of~\eqref{eq:Linear_evolution_intruder} are negligible, as shown in Fig.~\ref{fig:MeanEnergyVsTime}.

\section{Discussion}  
From a theoretical point of view, the considered model provides additional elements to the long-standing debate about the equilibrium nature of negative temperature states. To the best of our knowledge, this is the first time that an $H$-theorem is rigorously proven for this kind of systems, possibly supporting the generalization of the Second Law to negative temperature states originally proposed by Ramsey~\cite{Ramsey1956}. The necessity -- and even the legitimacy -- of an equilibrium description involving negative temperature was repeatedly questioned in the past~\cite{Dunkel2013, romero13, Struchtrup2018, calabrese18}: the fact that in this case the steady state verifies detailed balance seems to exclude any doubt about its equilibrium nature, providing therefore an important counterexample. 

\new{Furthermore, detailed balance also guarantees that the system can genuinely be considered as an {\it equilibrium} thermal bath acting at both positive and negative temperatures. The discussion in Sec.~\ref{sec:thermal_bath} provides clear evidence of the effectiveness of this approach in a simple yet non-trivial example.
Thus, the thermalization properties of the dynamics may reveal useful in designing numerical thermal baths at negative temperature in more realistic situations.}

At the same time, the model studied here is expected to be relevant for practical applications concerning complex conservative dynamics with bounds, encountered in different fields of physics. Equation~\eqref{eq:evolution_two_particle_neg} can be used for instance to reproduce the evolution of a system of isolated nuclear spins in the presence of high external field: the presented $H$-theorem ensures convergence to equilibrium, hence providing a handy, physically meaningful alternative to Monte Carlo simulations. 

Our results may also prove relevant in all those contexts where random exchange models are already used, but bounds are usually not taken into account.
 Kinetic models for flocks of birds such as those studied in~\cite{carrillo2010asymptotic,cao2020asymptotic}, for instance, are based on the unbounded version of Eq.~\eqref{eq:evolution_two_particle_neg}, where the bird velocities play the role of the energies. In this case it may be reasonable to assume bounds on the individual velocities (too fast birds would risk to leave the flock). Future research work may extend the present results to the kinetic models for flocks, and compare the stationary  velocity distributions with the actual observations, for which many experimental data are available~\cite{Cavagna2010, Cavagna2022}.
Another example is represented by the models for money exchange, largely used to describe and predict wealth distribution~\cite{greenberg2023twenty}. In this case it could be interesting to introduce ``soft'' upper bounds on the individual wealth that reduce, without suppressing, the possibility of large concentration of money. This correction may account for the combined effect of progressive taxation and market regulation. To have an idea of the striking effect of bounds, one can consider the extreme case study where the wealth of the single agent cannot exceed a given threshold $\Ener_M$, as in the dynamics~\eqref{eq:evolution_two_particle_neg}. By imposing $\Ener_M=2\Ener_T/N$, where $N$ is the number of agents and $\Ener_T$ is the total amount of money, the system reaches a steady state with uniform distribution ($\beta=0$): this is clearly impossible in the unbounded version of the model, where the stationary pdf is always exponential with $\beta>0$.
As an additional future perspective let us notice that, in many cases, imposing the same threshold for all the particles may reveal unrealistic. In analogy with~\cite{chatterjee2003money}, where authors consider heterogeneous saving properties, it could be therefore interesting to take bounds distributed according to a given pdf, in order to consider the natural variability of the social context.

\section*{Acknowledgments}
DL and MB were supported by ERC Advanced Grant RG.BIO (Contract No. 785932)    
\section*{References}
\bibliographystyle{unsrt.bst}
\bibliography{biblio}
\end{document}